\documentclass{article}[12pt]
\usepackage[T1,T2A]{fontenc}
\usepackage{amsmath,amssymb}
\usepackage{graphicx}%
\usepackage{xcolor}
\usepackage{verbatim}
\usepackage{multirow}

\newtheorem{stat}{Statement}
\newtheorem{property}{Property}

\newcommand{\fig}[4]{%
\begin{center}
\parbox{#2cm}{%
\refstepcounter{figure}\includegraphics[width=#2cm,height=#3cm]{#1}\\ \noindent Fig. \thefigure:\quad
#4}\end{center}}

\newcounter{strochka}

\newcounter{spisok}
\newcommand{\Prop}[2]{%
\begin{property}\hspace{-4pt}\textbf{(#1).}\\
#2
\end{property}}
\begin{document}

\begin{center}
{\bf \Large Yu. G  Ignat'ev\footnote{Institute of Physics, Kazan Federal University, Research Laboratory of Cosmology, 420008 Russia, Kazan, st. Kremlevskaya 18; email: yurii.ignatev.1947@yandex.ru} }\\[12pt]
{\bf \Large Self-gravitating Higgs field of scalar charge} \\[12pt]
\end{center}

\abstract{The self-gravitating Higgs field of a scalar charge has been studied. It is shown that in the zero and first approximation of the smallness of the scalar charge, the gravitational field of the scalar charge is described by the Schwarzschild-de Sitter metric with a cosmological constant determined by the vacuum potential of the Higgs field. An equation for the perturbation of the vacuum potential is obtained and studied. Particular exact solutions of the field equation are given. It is shown that in the case of a naked singularity, solutions to the field equation have the character of microscopic oscillations with a Compton wavelength. Limiting asymptotic cases of the behavior of solutions are studied and their comparative analysis is carried out in relation to the Fisher solution. The averaging of microscopic oscillations of the scalar field was carried out and it was shown that at $\Lambda>0$ they make a negative contribution to the macroscopic energy of the scalar field, reducing the observed value of the Black Hole mass. A computer simulation of a scalar field has been carried out, demonstrating various types of behavior of solutions.\\

}


%
\section{Introduction}
Since scalar-gravitational instability of the cosmological medium of scalarly charged fermions apparently results in the formation of scalarly charged black holes \cite{Yu_GC_23_No4} -- \cite{archive3}, it is necessary to consider in more detail the issue of such isolated static black holes.

The Lagrange function $L_s$ of the scalar Higgs field is\footnote{Here and below, Latin letters run through the values $\overline{1,4}$, Greek letters -- $\overline{1,3}$. The Planck system of units $G=c=\hbar=1$ is used throughout.\label{Plank_units}}
\begin{eqnarray} \label{L_s}
L_s=\frac{1}{16\pi}(g^{ik} \Phi_{,i} \Phi_{,k} -2V(\Phi)),
\end{eqnarray}
where
\begin{eqnarray}
\label{Higgs}
V(\Phi)=-\frac{\alpha}{4} \left(\Phi^{2} -\frac{m_s^{2}}{\alpha}\right)^{2}
\end{eqnarray}
-- potential energy of the scalar field, $\alpha$ -- self-action constant, $m_s$ -- boson mass.
The energy tensor - momentum of scalar fields relative to the Lagrange function \eqref{L_s} is:
\begin{eqnarray}\label{T_s}
T^i_{k} =\frac{1}{16\pi}\bigl(2\Phi^{,i}\Phi_{,k}- \delta^i_k\Phi_{,j} \Phi^{,j}+2V(\Phi)\delta^i_k \bigr),
\end{eqnarray}

Einstein's equations look like:
\begin{equation}\label{Eq_Einst_G}
R^i_k-\frac{1}{2}\delta^i_k R=8\pi T^i_k + \delta^i_k \Lambda_0,
\end{equation}
where $\Lambda_0$ is the initial value of the cosmological constant, associated with its observed value $\Lambda$, obtained by removing the constant terms in the potential energy, by the relation:
\begin{equation}\label{lambda0->Lambda}
\Lambda=\Lambda_0-\frac{1}{4}\frac{m_s^4}{\alpha}.
\end{equation}

In curvature coordinates (see, for example, \cite{Land_Field})
\begin{equation}\label{metric_stat}
ds^2=\mathrm{e}^{\nu(r)}dt^2-\mathrm{e}^{\lambda(r)}dr^2-r^2 d\Omega^.
\end{equation}
\begin{eqnarray}\label{T^i_k}
T^1_1=-\frac{\mathrm{e}^{-\lambda(r)}}{16\pi}{\Phi'}^2-\frac{\alpha}{32\pi}\left( \Phi^2-\frac{m^2_s}{\alpha}\right)^2,\quad (\equiv -p_\parallel)\nonumber\\
T^2_2=T^3_3=T^4_4=\frac{\mathrm{e}^{-\lambda(r)}}{16\pi}{\Phi'}^2-\frac{\alpha}{ 32\pi}\left(\Phi^2-\frac{m^2_s}{\alpha}\right)^2,\quad (\equiv -p_\perp=\varepsilon),
\end{eqnarray}
where $p_\parallel$ is the radial pressure, $p_\perp$ is the pressure along the surface of the sphere, $\varepsilon$ is the energy density of the scalar field.

\section{Massless scalar field - Fisher's solution}
\subsection{Fisher solutions}
For the first time, the metric of a scalarly charged black hole in the case of a massless canonical scalar field was found in the work of I.Z. Fisher (1948) \cite{Fisher}. Let us briefly present the main results of this work needed here.
In the curvature coordinates \eqref{metric_stat} and \emph{in the case of a massless scalar field} using the known first integral of the massless scalar field equation (the prime denotes the derivative with respect to the radial variable $r$):
\begin{equation}\label{Phi'=}
\Phi'=\displaystyle -\frac{G}{r^2}\mathrm{e}^{\frac{\lambda-\nu}{2}},
\end{equation}
where $G$ is the singular scalar charge, Fisher reduced Einstein's two independent equations to the following\footnote{Since \cite{Fisher} contains some mathematical inaccuracies, in order to avoid confusion we will briefly reproduce the results of this work, rewriting these equations in our notation and slightly reformatting Fisher's solution.}:
\begin{eqnarray}\label{Eq_Einst_Fish}
\mathrm{e}^{-\lambda}(1+r\nu')-1=-{\Phi'}^2r^2 \mathrm{e}^{-\lambda}; & \mathrm{e}^{-\lambda}(1-r\lambda')-1={\Phi'}^2r^2 \mathrm{e}^{-\lambda}.
\end{eqnarray}
The sum of these equations can be represented as:
\begin{equation}\label{sum_einst_fish}
(r^2e^{\nu-\lambda})'=2re^{-\nu}.
\end{equation}
Next Fisher using the function
\begin{equation}\label{Z->W}
W(r)=\displaystyle r\mathrm{e}^{\frac{\nu-\lambda}{2}}
\end{equation}
determines the solutions of the Einstein - Laplace equations system:
\begin{eqnarray}\label{resolve}
\mathrm{e}^\nu=\frac{1}{r}WW';\;\mathrm{e}^\lambda=r\frac{W'}{W};\;\Phi'=-\frac{G}{r}W.
\end{eqnarray}
Substituting \eqref{resolve} into the equations \eqref{Phi'=}, \eqref{Eq_Einst_Fish}, \eqref{sum_einst_fish} leads to a second-order closed differential equation for the function $W(r)$:, with the help of which it is easy to find the first integral, in turn, leading to a first-order equation with separable variables:
\begin{eqnarray}\label{1_integral_fisher}
rW'-W+\frac{a^2}{W}=C_1\equiv 2km \Rightarrow \frac{WW'}{W^2+2kmW-a^2}=\frac{1}{r},\quad (a^2\equiv kG^2).
\end{eqnarray}
where $C_1,k$ are arbitrary integration constants, $m$ is the singular mass. Thus, the problem is solved in quadratures, the study of the solution is a matter of technology. Relations \eqref{resolve} -- \eqref{1_integral_fisher} we
and will be called \emph{Fisher solutions}.

\subsection{Properties of Fisher solutions}
Next, we will deviate somewhat from the cited work of Fisher, introducing a new dimensionless function $f(r)$
\begin{eqnarray}\label{W-F}
W(r)\equiv \kappa(f(r)-p), & \displaystyle f(r)\geqslant p\equiv \frac{km}{\sqrt{k^2m^2+a^2}}\ equiv\frac{km}{\kappa},& \kappa\equiv \sqrt{k^2m^2+a^2},
\end{eqnarray}
with the help of which the solution to the equation \eqref{1_integral_fisher} can be written in the form of an algebraic equation for the function:
\begin{eqnarray}\label{Eq_F}
|f^2-1|^{1/2} \left|\frac{f+1}{f-1}\right|^p=\frac{C_2 r}{\sqrt{k^2m^2+ a^2}},
\end{eqnarray}
where $C_2$ is the integration constant.

Assuming that at infinity the metric \eqref{metric_stat} tends to pseudo-Euclidean, i.e.,
\begin{equation}\label{nu_lambda->0}
\left.\nu(r)\right|_{r\to\infty}\to 0,\; \left.\lambda(r)\right|_{r\to\infty}\to 0,
\end{equation}
we get from \eqref{Z->W}
\begin{equation}\label{W(8)}
\left.W(r)\right|_{r\to\infty}\to r.
\end{equation}
But then according to \eqref{W-F}%
\begin{equation}\label{nu_lambda->0}
\left.f(r)\right|_{r\to\infty}=\frac{r}{\kappa}\to\infty.
\end{equation}
Comparing this expression with the equation \eqref{Eq_F} in the limit $r\to\infty$, we find $C_2=1$. Thus, we bring the equation \eqref{Eq_F} to its final form:
\begin{eqnarray}\label{Eq_f_1}
|f^2-1|^{1/2} \left|\frac{f+1}{f-1}\right|^p=\xi,\quad \left(\xi\equiv\frac{r }{\kappa};\;\kappa=\sqrt{k^2m^2+kG^2}\right).
\end{eqnarray}
The equation \eqref{Eq_f_1} defines a one-parameter family of solutions $f(x;p)$, which, in turn, using the formulas \eqref{resolve} completely determines the solution to the problem:
\begin{eqnarray}\label{f(x;p)}
  W=\kappa(f-p); \; \mathrm{e}^\nu=\frac{f'_\xi}{\xi}(f-p);\; \mathrm{e}^\lambda=\frac{\xi f'_\xi}{f-p};\; \Phi'_\xi=-G\frac{f'_\xi}{\xi}.
\end{eqnarray}
Thus, the metric is also determined by a one-parameter family of functions, while the scalar potential $\Phi$ depends on two parameters, $\kappa,p$, but its derivative $\Phi'$ is still determined by a one-parameter family of functions.

Let us indicate exact solutions of the equation \eqref{Eq_f_1} for particular values of the parameter $p=0.1/2$.
\begin{eqnarray}
p=0:& \Rightarrow f=\sqrt{1+\xi^2},\; e^\nu=1, \nonumber\\
e^\lambda=&\displaystyle \frac{\xi^2}{1+\xi^2};\;\Phi'_\xi=\kappa G\frac{\sqrt{1+\xi^2} }{\xi};\nonumber\\
\Phi=& \displaystyle \kappa G\left(\sqrt{1+\xi^2}-\ln\frac{\sqrt{1+\xi^2}+1}{\sqrt{1+\xi^ 2}-1}\right)
\end{eqnarray}
-- in this case $m=0$ and the metric is generated by the massless charge $G$.
\begin{eqnarray}
p=\frac{1}{2}: \quad \displaystyle \Rightarrow f=\xi-1, e^\nu=1-\frac{3}{2\xi},\nonumber\\
e^\lambda= \displaystyle \left(1-\frac{3}{2\xi}\right)^{-1};\;\Phi'_\xi=\kappa G\left(1-\frac {3}{2\xi}\right);\;\Phi= \displaystyle \kappa G\left(\xi-\frac{3}{2}\ln \xi\right)
\end{eqnarray}
-- in this degenerate case $m^2=G^2/3k$, $\kappa=2km$, and the metric, up to renotations, coincides with the Schwarzschild metric. In both of these cases, the scalar potential has asymptotics
\begin{eqnarray}
\left.\Phi(\xi)\right|_{\xi\to0}\sim \ln \xi; & \displaystyle \left.\Phi'(\xi)\right|_{\xi\to0}\sim \frac{1}{\xi};\nonumber\\
\left.\Phi(\xi)\right|_{\xi\to\infty}\sim \kappa G\xi; & \displaystyle \left.\Phi'(\xi)\right|_{\xi\to\infty}\sim \kappa G=\mathrm{Const}.\nonumber
\end{eqnarray}
?
Solutions with a massless scalar field \emph{in other coordinate systems} are studied in detail in the work \cite{bronnik_fabris}, (see also reviews \cite{bronnik_rus} -- \cite{bronnik_eng}).

\section{Scalar field with the Higgs potential of a point scalar charge in the pseudo-Euclidean metric}
Let us now study the gravitational field generated by a scalar charge with the Higgs potential in the metric \eqref{metric_stat}. The scalar Higgs field equation $\Phi(r)$ in this metric has the form:
\begin{eqnarray}\label{Eq_C}
\displaystyle \frac{1}{r^2}\frac{d}{dr}\left(r^2\mathrm{e}^{\frac{\nu-\lambda}{2}}\frac{d }{dr}\Phi\right)-\mathrm{e}^{\frac{\nu-\lambda}{2}}\Phi(m^2_s-\alpha\Phi^2)=0.
\end{eqnarray}

In the work \cite{Yu_Scalar} the equation \eqref{Eq_C} was solved in the pseudo-Euclidean metric $\nu=\lambda=0$ for the central point scalar charge $G$. In this case, with the self-action constant $\alpha=0$, the equation
\eqref{Eq_C} reduces to the well-known Yukawa equation
\begin{eqnarray}\label{EqYukava}
\displaystyle \frac{1}{r^2}\frac{d}{dr}\left(r^2\frac{d}{dr}\Phi\right)-m_s^2\Phi=0
\end{eqnarray}
and has as its solution the well-known Yukawa potential
\begin{equation}\label{Yukava}
\Phi=\displaystyle{\frac{2G}{r}\mathrm{e}^{-m_sr}},
\end{equation}
where $G$ is a scalar charge.

The self-action constant factor $\alpha\not\equiv0$ fundamentally changes the nature of solutions to the equation \eqref{Eq_C}. Now this equation has no stable solutions with zero asymptotics at infinity
\begin{equation}\label{Phi(8)=0}
\left.\Phi(r)\right|_{r\to\infty}\to 0.
\end{equation}
Stable solutions of the equation \eqref{Eq_C} in a spatially flat metric are solutions with non-zero asymptotic behavior at infinity corresponding to special stable points of the dynamical system, --
\begin{equation}\label{Phi(8)=Phi0}
\left.\Phi(r)\right|_{r\to\infty}\to \Phi_\pm=\pm \frac{m_s}{\sqrt{\alpha}}.
\end{equation}
For solutions close to stable, assuming
\begin{equation}\label{Phi0+psi}
\Phi(r)=\Phi_\pm+\phi(r),\quad (\phi\ll 1),
\end{equation}
in the linear approximation we obtain the equation instead of \eqref{EqYukava}
\begin{equation}\label{Eq_phi}
\frac{1}{r^2}\frac{d}{dr}\biggl(r^2\frac{d\phi}{dr}\biggr)+2m_s^2\phi=0.
\end{equation}
Let us pay attention to the change in sign of the massive term compared to the Yukawa equation \eqref{EqYukava}, due to which the stable solution of the equation for the Higgs field will be \cite{Yu_Scalar}
\begin{equation}\label{Phi+phi=}
\Phi(r)=\pm\frac{m_s}{\sqrt{\alpha}}+\frac{C_1}{r}\cos(\sqrt{2}m_sr)+\frac{C_2}{r}\ sin(\sqrt{2}m_sr)\Rightarrow \pm\frac{m_s}{\sqrt{\alpha}}+\frac{2G}{r}\cos(\sqrt{2}m_sr)
\end{equation}
-- instead of exponential decay of the Yukawa potential \eqref{Yukava} we have a quasiperiodic potential.

The presence of a \emph{fundamental} scalar field with the Higgs potential fundamentally changes the physical picture. Now the vacuum state corresponds to one of the stable points of the Higgs potential \eqref{Phi(8)=Phi0}, which, in turn, corresponds to the zero energy of the scalar field.
\section{Self-gravitating scalar field with Higgs potential}
\subsection{Field equations}
Taking into account the above, we study the solution to the complete problem of a self-gravitating scalar Higgs field. Nontrivial combinations of Einstein's equations with a cosmological constant in the metric \eqref{metric_stat}\footnote{These are combinations of the equations $^1_1$, $^4_4$ and the scalar field equation.} can be reduced to the form:
\begin{eqnarray}\label{Eq_A}
r\Phi'^2+(\lambda+\nu)'=0;\\
\label{Eq_B}
\!\!\!\!\mathrm{e}^\lambda-1-r\nu'-\!r^2\mathrm{e}^\lambda \left[\Lambda-\frac{\alpha}{2}\left(\Phi^2-\frac{m^2_s}{\alpha}\right)^2\right]=0.
\end{eqnarray}

We will look for solutions to the system of equations \eqref{Eq_C}, \eqref{Eq_A}, \eqref{Eq_B} that are close to stable, assuming \eqref{Phi0+psi}. Then, in the zero approximation, due to the smallness of $\phi(r)$, the equation \eqref{Eq_C} becomes an identity, and the equation \eqref{Eq_A} gives
\begin{equation}\label{lambda=-nu}
\lambda=-\nu.
\end{equation}
As a result, the equation \eqref{Eq_B} will be reduced to a closed-loop equation for $\nu$ (or $\lambda$)
\begin{eqnarray}
\label{Eq_B0}
r\nu'+1+\mathrm{e}^{-\nu}(1-\Lambda r^2)=0,
\end{eqnarray}
solving which, we find:
\begin{equation}\label{nu0}
\nu_0=-\lambda_0=\ln\left(1-\frac{2m}{r}-\frac{\Lambda r^2}{3}\right),
\end{equation}
where $m$ is the constant of integration. Thus, in the zeroth approximation we obtain the well-known Schwarzschild-de Sitter solution \cite{Edd}:
\begin{eqnarray}\label{Shvarc-deSit}
ds^2= \displaystyle \left(1-\frac{2m}{r}-\frac{\Lambda r^2}{3}\right)dt^2 -\left(1-\frac{2m}{ r}-\frac{\Lambda r^2}{3}\right)^{-1}dr^2-r^2d\Omega^2.
\end{eqnarray}

At first glance, it seems that the solution \eqref{Shvarc-deSit} does not depend on the scalar field $\Phi(r)$. However, it is not. To correctly interpret this solution, we must first take into account the formula \eqref{Phi(8)=Phi0} for
unperturbed scalar field $\Phi_\pm$ and, secondly, a formula for renormalizing the observed cosmological constant \eqref{lambda0->Lambda}, putting in \eqref{Shvarc-deSit}
\[\Lambda=\Lambda_0-\frac{1}{4}m^2_s\Phi^2_\pm.\]
Thus, the solution \eqref{Shvarc-deSit}, in addition to the central mass $m$, is also determined by the square of the unperturbed value of the scalar potential, i.e., ultimately, by the square of the scalar charge.

It turns out that in the first approximation of the smallness of $\phi$ the solution \eqref{Shvarc-deSit} remains valid. Indeed, due to \eqref{Eq_A}, in the first approximation the relation \eqref{lambda=-nu} is preserved, and therefore the equation \eqref{Eq_B0} is also preserved. Thus, the \emph{ metric \eqref{Shvarc-deSit} is preserved in the approximation linear in $\phi$}. Therefore, in the linear approximation, the field equation \eqref{Eq_C} can be considered against the background of the Schwarzschild - de Sitter solution \eqref{Shvarc-deSit}. So, in the linear approximation \eqref{Phi0+psi} we obtain the equation for the perturbation of the scalar field $\phi(x)$
\begin{eqnarray}\label{Eq_C20}
\displaystyle \frac{d^2\phi}{dr^2}+\frac{d}{dr}\ln\bigl(r^2\mathrm{e}^{\nu_0(r)}\bigr)\frac{d\phi}{dr}+2m^2_s\phi=0.
\end{eqnarray}
Introducing the dimensionless variable $x$ and dimensionless non-negative parameters $\gamma,\sigma$
\begin{eqnarray}\label{new_var}
x= \frac{r}{2m};\; \gamma=\frac{4}{3}\Lambda m^2\geqslant0;\; \sigma=2\sqrt{2}mm_s\geqslant0,
\end{eqnarray}
Let's rewrite the equation \eqref{Eq_C20} in terms of these quantities:
\begin{eqnarray}\label{Eq_C2}
\displaystyle \frac{d^2\phi}{dx^2}+\frac{d}{dx}\ln\bigl(x(x-1-\gamma x^3)\bigr)\frac{d\phi}{dx}+\sigma^2\phi=0& \Rightarrow\nonumber\\
\displaystyle \frac{d^2\phi}{dx^2}+\frac{1-2x+4\gamma x^3}{x(1-x+\gamma x^3)}\frac{d\phi}{dx}+\sigma^2\phi=0.
\end{eqnarray}

For convenience of analysis, as well as numerical integration, we will consider the second-order linear homogeneous differential equation \eqref{Eq_C2} also in the form of a normal system of first-order equations:
\begin{eqnarray}\label{Eq_Sys}
\frac{d\phi}{dx}=z(x);\quad \displaystyle \frac{dz}{dx}=-\frac{1-2x+4\gamma x^3}{x(1-x+\gamma x^3)} z-\sigma^2\phi.
\end{eqnarray}

\subsection{Horizons and singularity}
Solutions of the field equation $\phi(x)$ \eqref{Eq_C2} are largely determined by the horizons and singularity of the Black Hole, which determine the behavior of the argument of the logarithmic function in this equation
\begin{equation}\label{r^2U=0}
r^2\mathrm{e}^{\nu_0}\equiv r^2\biggl(1-\frac{2m}{r}-\frac{\Lambda r^2}{3}\biggr)=0\Rightarrow x (1-x+\gamma x^3)=0.
\end{equation}
The singularity corresponds to the zero root of the equation \eqref{r^2U=0} $x_0=0$, and the horizons, if they exist, correspond to the positive real roots of the cubic equation
\begin{equation}\label{x_i}
\mathrm{e}^{\nu_0}=0\Rightarrow \gamma x^3-x+1=0.
\end{equation}

The discriminant of the cubic equation \eqref{x_i}, $\Delta$, is equal to:
\begin{equation}\label{Delta}
\Delta=\gamma(4-27\gamma).
\end{equation}
For $\Delta>0$ all roots of the horizon equation \eqref{x_i} are real, for $\Delta<0$ one root is real and two are complex conjugate, $\Delta=0$ -- all three roots are real and distinct, and, at least two of them are the same. At
\begin{equation}\label{gamma<4/27}
\gamma>0.
\end{equation}
and $\gamma<4/27$ $\Delta>0$ all three roots are real, and one of them, $x_0<-3$, is negative and two are positive: $1<x_1<\frac{3} {2}$, $x_2>\frac{3}{2}$. Thus, for $0<\gamma<4/27$, the metric has two horizons: $r_1$ is internal and $r_2$ is external:
\begin{equation}\label{r_hor}
2m<r_1<3m, \quad r_2>3m.
\end{equation}
At $\gamma=4/27$ both horizons merge into one doubly degenerate $r_1=r_2$. At $\gamma>4/27$ there are no horizons, and the gravitational field of the Black Hole is described by a metric with a naked singularity $r=0$. At $\gamma\equiv0$
only the Schwarzschild horizon remains. For $\gamma<0$ and $\Delta<0$, -- in this case there is also one real positive root, which corresponds to one horizon $x_1<1$.\footnote{Note that according to the cosmological constant renormalization formula \eqref {lambda0->Lambda} we have no right to discard the case $\gamma<0$ from consideration.}

Thus, depending on the value of $\gamma$, three-dimensional space is divided by horizons into $\mathbf{R}$ - and $\mathbf{T}$ - regions along the radial variable $x$
\begin{eqnarray}\label{R-T}
\begin{array}{llll}
\gamma\leqslant 0:  & \mathbf{X_1}=[0,x_1], (\mathrm{T}); &  \mathbf{ X_3}=(x_1,+\infty), (\mathrm{R}); & \\[12pt]
\displaystyle 0<\gamma<\frac{4}{27}: & \mathbf{X_1}=[0,x_1], (\mathrm{T}); &  \mathbf{ X_2}=(x_1,x_2), (\mathrm{R}); & \mathbf{X_3}=(x_2,+\infty), (\mathrm{T});\\[12pt]
\displaystyle \gamma>\frac{4}{27}: & \mathbf{ X_3}=[0,+\infty), (\mathrm{T}).&   &\\
\end{array}
\end{eqnarray}
Below Figure \ref{Ignatev1} shows the behavior of solutions to the equation \eqref{Eq_C2} in the regions $\mathbf{X_1},\mathbf{X_2},\mathbf{X_3}$ in the case of $\gamma=0.1<4/27$ , corresponding to the initial values in each of the areas
\begin{eqnarray}
\mathbf{X_1:} \phi(0)=\pm1,z(0)=0; & \mathbf{X_2:} \phi(2)=\pm1,z(0)=0;& \mathbf{X_3:} \phi(3)=\pm1,z(0)=0.\nonumber
\end{eqnarray}
In this case, we assumed $\sigma=1$.

\subsection{Specific solutions}
In two special cases, the equation \eqref{Eq_C2} is solved in quadratures.
\subsubsection{Massless scalar field $m_s=0$}
Note that, strictly speaking, we have no right to consider the case of zero mass of scalar bosons, since at $m_s=0$ the scalar Higgs potential \eqref{Higgs} degenerates into a parabolic potential, i.e., in this case we go beyond the scope of the study models. Stable points of the dynamic system \eqref{Phi(8)=Phi0} $\Phi_\pm$ degenerate into one zero point $\Phi_+=\Phi_-=0$. Only this trivial solution $\Phi=0$ is now stable.
Therefore, within the framework of our model, we can only consider an asymptotically massless scalar field in the sense of approximation:
\begin{equation}\label{m_sr->0}
m_s r\to 0,
\end{equation}
i.e., in the region $r\to0$. In this case, the equation \eqref{Eq_C2} is immediately integrated
\begin{equation}\label{phi_m_s=0}
\phi=C_1 +C_2\int\frac{dx}{x(\gamma x^3-x+1)}\equiv C_1+C_2 J(x),
\end{equation}
Where
\begin{eqnarray}
\label{J(x)}
J(x)= \displaystyle \int\frac{dx}{x(\gamma x^3-x+1)}.
\end{eqnarray}
In particular, for $\Lambda=0\Rightarrow\gamma=0$ this integral is easily calculated
\begin{equation}\label{phi_m_s=Lambda=0}
\phi(x)=C_1+C_2\ln\biggl|\frac{x-1}{x}\biggr|\Rightarrow \Phi=\frac{m_s}{\sqrt{\alpha}}+C_2\ln\left|1-\frac{2m}{r}\right|, \; (m_s\to0, \Lambda=0)
\end{equation}
and gives the asymptotics at infinity
\begin{eqnarray}\label{phi(8)_m_s=Lambda=0}
\left.\Phi(r)\right|_{r\to\infty}\backsimeq\frac{m_s}{\sqrt{\alpha}}-\frac{2m C_2}{r}.
\end{eqnarray}

For $\gamma\not=0$ the integral in \eqref{phi_m_s=0} is also calculated in elementary functions
\begin{eqnarray}
\label{int_phi}
J(x)= \displaystyle \ln|x|+\sum\limits_{i=1}^{3}\delta_i\ln|x-x_i|;\quad
\delta_i\equiv \frac{1-\gamma x^2_i}{3\gamma x^2_i-1},
\end{eqnarray}
where $x_i$ are the roots of the horizon surface equation
Thus, the solution \eqref{phi_m_s=0} in the case of non-degenerate roots $x_i$ \eqref{x_i} leads to logarithmic asymptotics at infinity:
\begin{equation}\label{asimpt_8_m_s=0}
\left.\Phi(r)\right|_{r\to\infty}\backsimeq\frac{m_s}{\sqrt{\alpha}}+C_2(1+\delta_1+\delta_2+\delta_3)\ln \frac {r}{2m}.
\end{equation}
%

In particular, the solution \eqref{phi_m_s=Lambda=0} is obtained from  \eqref{x_i} -- \eqref{int_phi} for $\gamma=0$, $x_1=1$, $\delta_1=-1$, in In this case, only one term is retained in the sum \eqref{int_phi}, corresponding to the simple horizon $x=x_1=1$.

However, due to the condition \eqref{m_sr->0}, for the correctness of this estimate it is necessary to satisfy the conditions
\begin{equation}\label{m<<l_s}
1\ll x\ll \frac{1}{2mm_s}\Rightarrow 2m\ll \frac{1}{m_s},
\end{equation}
i.e., the Compton wavelength of the scalar boson must be much larger than the horizon radius of the black hole and, in addition, $r_\infty\ll m^{-1}_s$.

\subsubsection{Zero cosmological constant $\mathbf{\Lambda\equiv0}$}
In this case, the solution to the equation \eqref{Eq_C2} is expressed through the confluent functions Heun,\\
$\mathrm{H}_c(2i\sigma,0,0,0,0,x)$, \cite{Heun}:
\begin{eqnarray}\label{psi_HeunC}
\phi(x) =\displaystyle C_1 \mathrm{e}^{2i\sigma x}\mathrm{H}_c(2i\sigma,0,0,0,0 x)+
C_2 \mathrm{e}^{2i \sigma x}\mathrm{H}_c(2i\sigma,0,0,0,0,x)\int \frac{\mathrm{e}^{-2i\sigma x}dx}{x(x - 1)\mathrm{H}^2_c(2i\sigma,0,0,0,0,x)}.
\end{eqnarray}
In the general case, functions $\mathrm{H}_c(x)$ have two regular and one irregular singularities of rank 1 at points $x=[0,1,\infty]$. In what follows, however, we will not use the exact solution \eqref{psi_HeunC}, taking into account, firstly, its particular nature, and, secondly, the fact that, unfortunately, the functions $\mathrm{HeunC}(x )$, which determine its solution for $\Lambda=0$, are still very unreliably tabulated in applied mathematical packages for sufficiently large arguments $x$. Therefore, we will directly integrate the equation \eqref{Eq_C2} using numerical methods.

\section{Asymptotic behavior of solutions to the equation \eqref{Eq_C2}}
\subsection{Behavior of solutions near the singularity $r=0$}
For $x\to0$ the field equation \eqref{Eq_C2} reduces to a simple second order differential equation
\begin{eqnarray}
\phi'' +\frac{\phi'}{x}+\sigma^2\phi=0,& x\to0,
\end{eqnarray}
which has its decisions
\begin{eqnarray}\label{phi(0)}
\phi(x)= C_1 \mathrm{I}_0(\sigma x)+C_2\mathrm{Y}_0(\sigma x)\backsimeq C_1+C_2\frac{2}{\pi}\ln \sigma x ,& (\sigma x \to 0),
\end{eqnarray}
where $\mathrm{I}_0(z)$ and $\mathrm{Y}_0(z)$ are Bessel functions of the 1st and 2nd kind, respectively. Thus, the scalar field potential diverges logarithmically near the singularity, and its derivative is equal to
\begin{equation}\label{phi'(0)}
\left.\Phi'\right|_{x\to0}\backsimeq \frac{C_1}{x}=\frac{G}{r}.
\end{equation}
\subsection{Behavior of solutions near horizons}
It is obvious that the solution of the system \eqref{Eq_Sys} near the horizons is singular, therefore the main term on the right side of the second equation \eqref{Eq_Sys} near the horizons is the term proportional to $z(x)$. Discarding the last term on the right side of this equation near the horizon $x_a$, we find by integrating:
\begin{eqnarray}
\left.z(x)\right|_{x\to x_a}\backsimeq C_1\frac{x}{\gamma x^3-x+1}.
\end{eqnarray}
Considering that $x_i$ are the roots of the horizon equation \eqref{r_hor}, we write according to Vieta's theorem
\[\gamma x^3-x+1=\gamma(x-x_a)(x-x_b)(x-x_c),\]
where, for definiteness, $x_b\not= x_a$ is a positive root of the equation \eqref{r_hor}, and $x_c$ is negative. Thus, near the horizon $x=x_a>0$ we obtain;
\[z(x)\backsimeq C_1\frac{x_a}{\gamma(x-x_a)(x_a-x_b)(x_a-x_c))}.\]
Integrating this relation, we obtain an asymptotic expression for the potential function $\psi(x)$ near the horizon $x=x_a$:
\begin{equation}\label{psi(x_a)}
\left.\phi(x)\right|_{x\to x_a}\sim \frac{C_1\ln|x-x_a|}{\gamma(x_a-x_b)(x_a-x_c)}+C_2.
\end{equation}
It is obvious that in the regions $\mathbf{X_1}$ and $\mathbf{X_2}$ the difference $x_a-x_b$ has opposite signs, which explains the discontinuity of the second kind in the function $\psi(x)$ when passing through the horizon. Behind the right horizon, the function $\psi(x)$ has the character of damped periodic oscillations.

\subsection{Behavior of the solution at infinity}
\subsubsection{Small values of $\gamma\ll 1$}
For small $\gamma$ in the intermediate range of values $x$
\begin{equation}\label{gamma<<1}
1\ll x\ll \frac{1}{\sqrt{\gamma}},\qquad (\gamma\ll 1)
\end{equation}
the equation \eqref{Eq_C2} reduces to a simple differential equation
\begin{equation}\label{EQ_x>>1}
\frac{d^2\psi}{dx^2}+\frac{2}{x}\frac{d\psi}{dx}+\sigma^2\psi=0,
\end{equation}
which has its solution
\begin{eqnarray}\label{sol_1/g>>x->8}
\phi(x)= c_1\frac{\sin\sigma x}{x}+c_2\frac{\cos\sigma x}{x}=\frac{C_1}{r}\sin\sqrt{2}m_s r+\frac{C_2}{r}\cos\sqrt{2}m_s r,
\end{eqnarray}
i.e., describes damped periodic oscillations with frequency $\omega$ and period $\tau$
\begin{equation}\label{Period}
\omega=\sigma,\; \tau=\frac{2\pi}{\sigma}\Rightarrow T=\frac{\sqrt{2}\pi}{m_s}.
\end{equation}
Thus, in the intermediate range of values of the radial variable $x$, up to redesignations $C_1=0,C_2=2G$, the solution to the field equation coincides with the solution in flat space-time \eqref{Phi+phi=}.
\subsubsection{$x\to\infty$, $\gamma x^2\gg 1$}
In area
\begin{equation}\label{x->8}
x\to\infty,\qquad \gamma x^2\gg 1
\end{equation}
the equation \eqref{Eq_C2} takes the form
\begin{equation}\label{EQ_x->8}
\frac{d^2\psi}{dx^2}+\frac{4}{x}\frac{d\psi}{dx}+\sigma^2\psi=0,
\end{equation}
and has its decision
\begin{eqnarray}\label{sol_x->8}
\phi(x)= \frac{C_1}{x^3}(\sigma x\cos{\sigma x}-\sin\sigma x)+\frac{C_2}{x^3}(\cos{\sigma x}+\sigma x \sin\sigma x).
\end{eqnarray}
And in this case we get damped oscillations with a period \eqref{Period}, however, the amplitude of the oscillations drops in proportion to $1/x^2$.

So, we note that in the case of small values of $\gamma$, an intermediate region can form with oscillations of the scalar field damping in proportion to $1/x$, which then quickly fall:
\begin{eqnarray}\label{1/x}
\gamma\ll 1,\qquad x\in \left(1,\frac{1}{\sqrt{\gamma}}\right): \displaystyle \phi\backsimeq \frac{\mathrm{e}^{i \sigma x}}{x}; \\
\label{1/x^2}
  \forall\gamma, \qquad \displaystyle x\in \left(\mathrm{Max}\left\{\frac{1}{\sqrt{\gamma}},1\right\},+\infty\right) : \displaystyle \phi\backsimeq \frac{\mathrm{e}^{i\sigma x}}{x^2}.
\end{eqnarray}
\section{Averaging of scalar potential oscillations}
The above analysis shows the oscillatory nature of the scalar field outside the horizon region. It should be emphasized that the oscillations of the scalar field have a purely microscopic character, corresponding to oscillations \emph{with the Compton wavelength } $\exp(i\sqrt{2}m_s r)$. In this case, a macroscopic observer can measure only some average dynamic quantities corresponding to these oscillations, in particular, macroscopic energy density and pressure. In this case, the macroscopic picture corresponds to some, generally speaking, anisotropic medium with macroscopic characteristics of pressure and energy density, as well as a macroscopic equation of state.
The situation here is completely analogous to microscopic oscillations of the scalar field at the late stages of the evolution of the Universe (see \cite{Yu_17} -- \cite{Yu_18}). The difference lies only in the nature of the oscillations - in the cosmological situation these are time oscillations $\exp(im_s t)$, in our situation they are spatial.

Expanding the expressions for the components of the energy-momentum tensor \eqref{T^i_k} in terms of the smallness of the perturbation $\phi$ of the scalar potential \eqref{Phi0+psi}, we obtain in the quadratic approximation:
\begin{eqnarray}
T^4_4=\varepsilon=\displaystyle \frac{\mathrm{e}^{\nu_0(r)}}{16\pi}{\phi'}^2-\frac{m^2_s}{8\pi }\phi^2;& \displaystyle -T^1_1=p_\parallel=\frac{\mathrm{e}^{\nu_0(r)}}{16\pi}{\phi'}^2+\frac {m^2_s}{8\pi}\phi^2.
\end{eqnarray}
Expressing these quantities through the variable $x$ and the functions $\phi(x)$ and $z(x)$ that we use, we obtain dimensionless expressions for the physical quantities $\varepsilon(r)$ and $p_\parallel(r)$
\begin{eqnarray}\label{p,e_psi}
16\pi(2m)^2\varepsilon=\displaystyle \mathrm{e}^{\nu_0(x)}z^2-\sigma^2\phi^2; & 16\pi(2m)^2 p_\parallel=\displaystyle \mathrm{e}^{\nu_0(x)}z^2+\sigma^2\phi^2.
\end{eqnarray}
where $\mathrm{e}^{\nu_0(x)}$ is described by the expression \eqref{x_i}.

Taking into account the rapidly oscillating nature of the solutions to the system of equations \eqref{Eq_Sys} for $\sigma x\gg 1$, let us average the values \eqref{p,e_psi} over a sufficiently large interval of the radial variable, using the technique of averaging cosmological fluctuations of the scalar field \cite{Yu_17} -- \cite{Yu_18}. Namely, let us introduce the macroscopic average of the rapidly varying function $f(r)$:
\begin{equation}\label{everage}
\overline{f(r)}= \frac{1}{T}\int\limits_{r-T/2}^{r+T/2}f(r)dr\Rightarrow \overline{f(x)}= \frac{1}{\tau}\int\limits_{x-\tau/2}^{x+\tau/2}f(x)dx,
\end{equation}
believing
\begin{equation}\label{WKB}
\tau x \gg 1.
\end{equation}
Assuming further that the asymptotic formulas are valid in the \eqref{WKB} approximation (see \eqref{1/x} -- \eqref{1/x^2})
\begin{equation}\label{phi-sim}
\phi(x)\backsimeq \phi_0 \mathrm{e}^{i\sigma x}\frac{1}{x^\beta}; \quad z(x)\backsimeq i\sigma\phi_0 \mathrm{e}^{i\sigma x}\frac{1}{x^\beta},
\end{equation}
we get
\begin{eqnarray}
\overline{\phi(x)}\approx 0; & \overline{z(x)}\approx 0;\nonumber\\
\overline{\phi^2(x)}\approx \frac{|\phi_0|^2}{x^{2\beta}}; & \displaystyle \overline{z^2(x)}\approx \frac{\sigma^2|\phi_0|^2}{x^{2\beta}}.\nonumber
\end{eqnarray}
Substituting these expressions into \eqref{p,e_psi}, we obtain for macroscopic average energy densities and pressures of scalar field oscillations:
\begin{eqnarray}\label{p,e_psi_average}
16\pi(2m)^2\ \overline{\varepsilon(x)}\backsimeq\displaystyle \bigl[\mathrm{e}^{\nu_0(x)}-1\bigr]\frac{\sigma^2 |\phi_0|^2}{x^{2\beta}}= -\frac{1+\gamma x^3}{x}\frac{\sigma^2|\phi_0|^2}{x^{ 2\beta}};\nonumber\\
16\pi(2m)^2\ \overline{p_\parallel(x)}\backsimeq\displaystyle -\bigl[\mathrm{e}^{\nu_0(x)}+1\bigr]\frac{\sigma ^2|\phi_0|^2}{x^{2\beta}}=-\frac{1-2x+\gamma x^3}{x}\frac{\sigma^2|\phi_0|^2}{ x^{2\beta}}.
\end{eqnarray}
Note, firstly, that the left-hand sides of the relations \eqref{p,e_psi_average} are expressions for the dimensionless normalized macroscopic average energy density and radial pressure of the scalar field. Secondly, we note that the oscillation energy density is negative $\varepsilon<0$. Further, in the region $\forall \gamma$ according to \eqref{1/x^2} we obtain
\[\left.\overline{\varepsilon}\ \right|_{\gamma x^3\to\infty}\backsimeq -\frac{\gamma\sigma^2|\phi_0|^2}{16\pi (2m)^2 x^2};\qquad \overline{p_\parallel}\to \overline{\varepsilon}.\]

Thus, in the region $\gamma x^3\gg1$, microscopic oscillations of the scalar field create a macroscopic background with a negative energy density and the equation of state $\overline{p}=-\overline{\varepsilon}$, i.e., they manifest themselves as a macroscopic phantom scalar field. Mass-energy of this field
\[M_s(r)=4\pi \int\limits_{r_0}^r \overline{\varepsilon} r^2dr \sim -\frac{\gamma m_s^2|\phi_0|^2}{8\pi }(r-r_0)\]
increases in proportion to the radius, thereby reducing the observed mass of the Black Hole.
Note, however, that according to \eqref{R-T} the $\forall \gamma$ region (see \eqref{1/x^2}) for $\gamma>0$ is a $\mathrm{T}$-region, in for which there is no stationary state and there is no infinitely distant observer. In fact, in $\mathrm{T}$-regions we must swap spatial and temporal coordinates. Therefore, for $\gamma>0$, the <<external region>> $\forall \gamma$ is, in fact, a cosmological continuation of the solution. Thus, to correctly interpret the static scalar charge field, we must set $\gamma<0$. In this case, a possible and correct interpretation of the formulas \eqref{p,e_psi_average} and those following them is that the macroscopic oscillation energy density becomes positive, and at the same time the macroscopic mass of the scalar halo also becomes positive. We intend to return to a more detailed study of this issue in the next part of the article.

\section{Numerical modeling}
\subsection{<<General>> numerical solutions}
To carry out numerical integration
Secondly, let us formulate an obvious property of the solutions of this system.

\Prop{General numerical solution}{\label{prop1}
Let $\mathbf{\Psi}_1(x0;x)$ and $\mathbf{\Psi}_2(x0;x)$ be solutions to the corresponding Cauchy problems for this system:
\begin{eqnarray}\label{X_1,X_2}
\mathbf{\Psi}_1(x;x_0)\equiv [\psi_1(x),z_1(x)]\mathbf{:}\quad [\psi_1(x_0)=1,z_1(x_0)=0]; \nonumber\\
\mathbf{\Psi}_2(x;x_0)\equiv [\psi_2(x),z_2(x)]\mathbf{:}\quad [\psi_2(x_0)=0,z_2(x_0)=1].
\end{eqnarray}
Then the solution to the Cauchy problem with arbitrary initial conditions
\begin{equation}\label{Coshe1}
\mathbf{\Psi}(x;x_0)\equiv [\psi(x),z(x)]\mathbf{:}\quad [\psi(x_0)=C_1,z(x_0)=C_2]
\end{equation}
There is:
\begin{equation}\label{Gen_Sol}
\mathbf{\Psi}(x;x_0)=C_1\mathbf{\Psi}_1(x;x_0)+C_2\mathbf{\Psi}_2(x;x_0).
\end{equation}

For convenience, we will call the solution \eqref{Gen_Sol} the general numerical solution of a system of linear homogeneous differential equations. $\blacksquare$
}

In what follows, we will use this property by default to study numerical models.

\subsection{Small values of the parameter $\gamma<4/27$}
In this case, as we noted above, the metric \eqref{Shvarc-deSit} has two horizons $H_\pm$ \eqref{r_hor}, through which it is impossible to analytically continue the solution of the field equations \eqref{Eq_Sys}. Figure \ref{Ignatev1} shows graphs of the potential function $\phi(x)$ for
\begin{eqnarray}\label{gamma=0.1}
\gamma=0.1<\frac{4}{27}\Rightarrow & x_1=1.153467305, & x_2= 2.423622140
\end{eqnarray}
in three areas
\[\mathbf{X_1}=[0,r_1);\quad \mathbf{ X_2}=(x_1,x_2);\quad \mathbf{X_3}=(x_2,+\infty).\]
In this case, the initial conditions were chosen as follows:
\[\mathbf{X_1}: x(1)=\pm1,\; x'(1)=0;\quad \mathbf{X_2}: x(1.2)=\pm1,\; x'(1.2)=0;\quad \mathbf{X_3}: x(3)=\pm1,\; x'(3)=0.\]
\fig{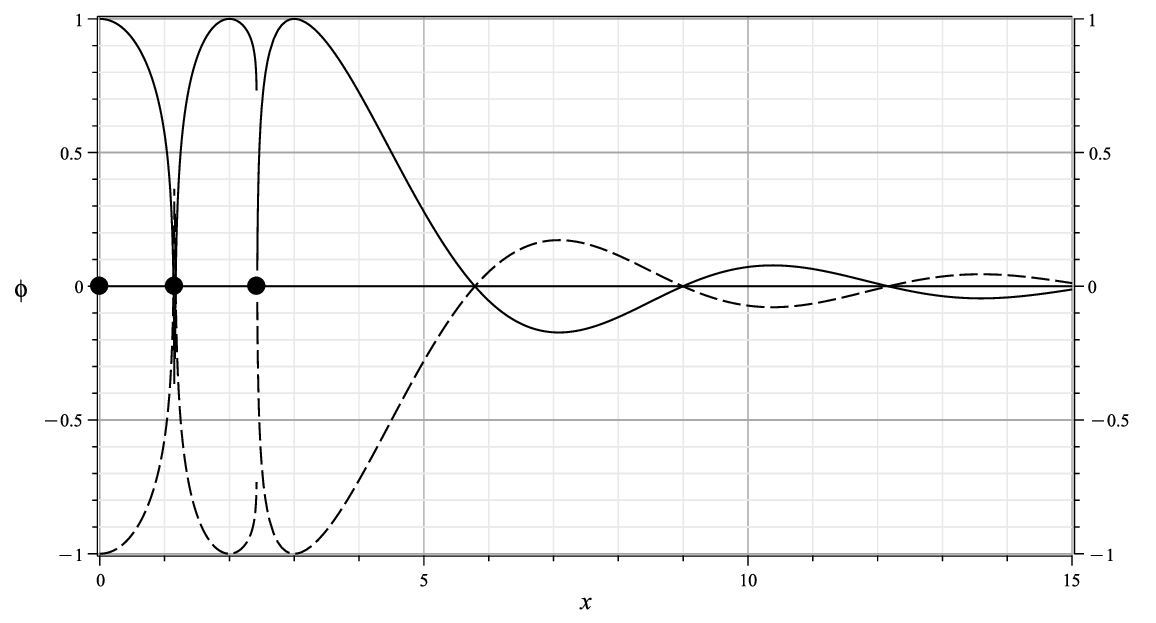}{12}{6}{\label{Ignatev1}Behavior of solutions to the equation \eqref{Eq_C2} in the case of $\gamma=0.1$ in the regions $\mathbf{X_1},\mathbf{X_2}, \mathbf{X_3}$. Solid lines correspond to positive initial values of $\phi$, dashed lines to negative ones. Black circles on the abscissa axis mark the radius of the singularity $x=0$ and the radii of the horizons $x_1\approx 1.153467305$ and $x_2\approx 2.423622140$.}

\subsection{Large values of the parameter $\gamma>4/27$}
At $\gamma>4/27$ there are no horizons in the \eqref{Shvarc-deSit} metric, i.e., we have a Black Hole with a bare singularity $r=0$. In this case, the field equations \eqref{Eq_Sys} have analytical solutions in the entire space $r\geqslant0$. Figure \ref{Ignatev2} and \ref{Ignatev3} show graphs of the potential function $\psi(x)$ and its derivative $\psi'(x)$ at
\begin{eqnarray}\label{gamma=0.2}
\gamma=0.2>\frac{4}{27}.
\end{eqnarray}
\fig{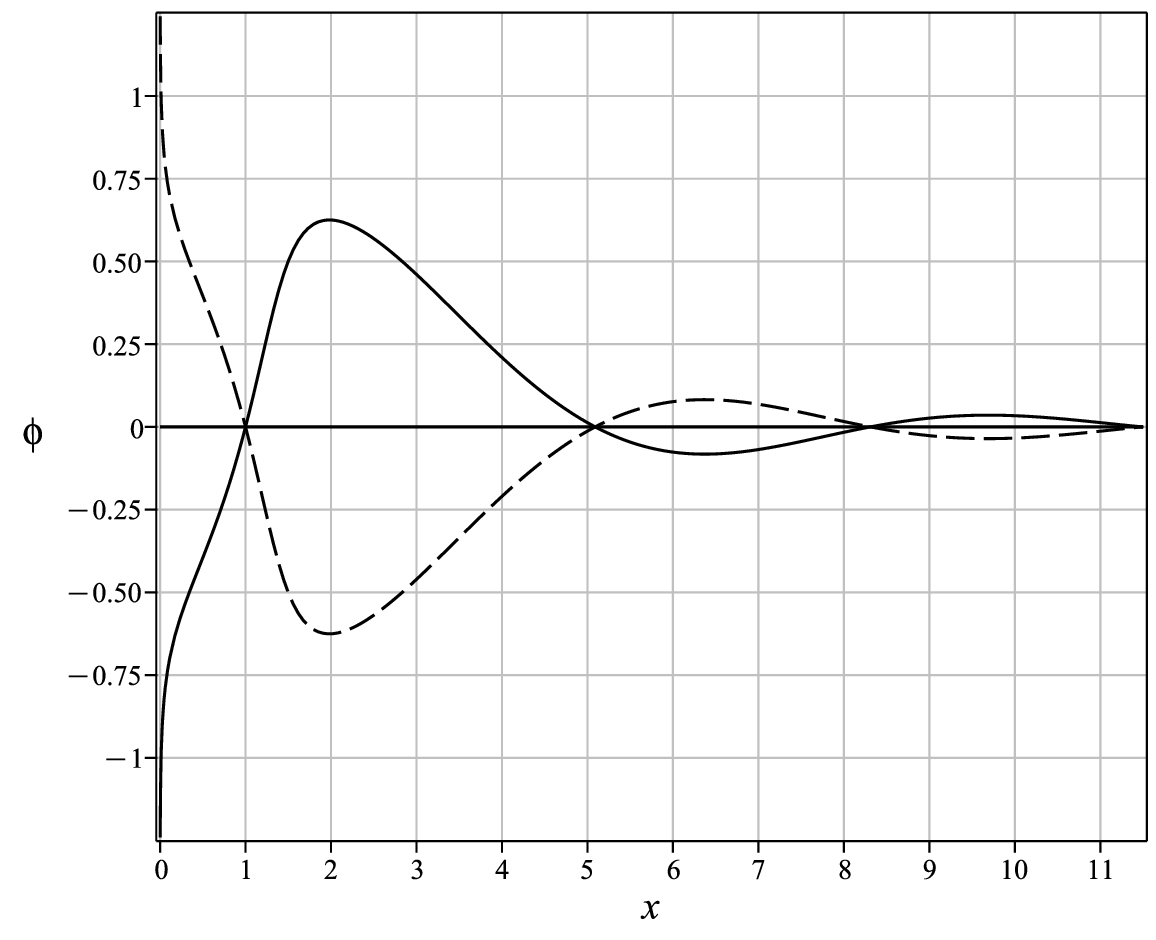}{10}{6.5}{\label{Ignatev2}Scalar potential function $\phi(x)$ for $\gamma=0.2$, $\sigma=1$.}
\fig{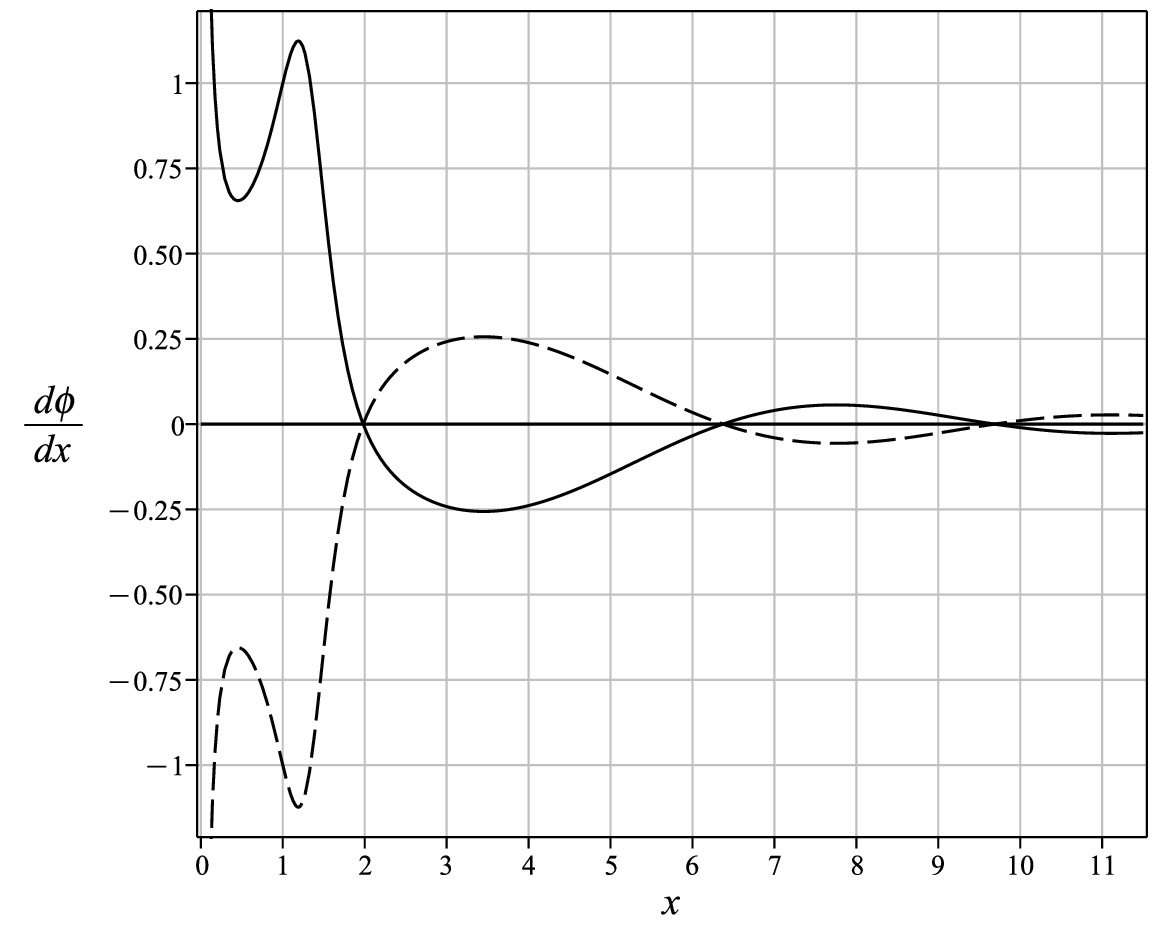}{10}{6.5}{\label{Ignatev3}Function of derivative of the scalar potential $\phi'(x)$ for $\gamma=0.2$, $\sigma=1$.}
As before, in the figures \ref{Ignatev2} - \ref{Ignatev3}, dashed lines indicate graphs of potential functions for negative values of the starting potential, and solid lines for positive ones.

\fig{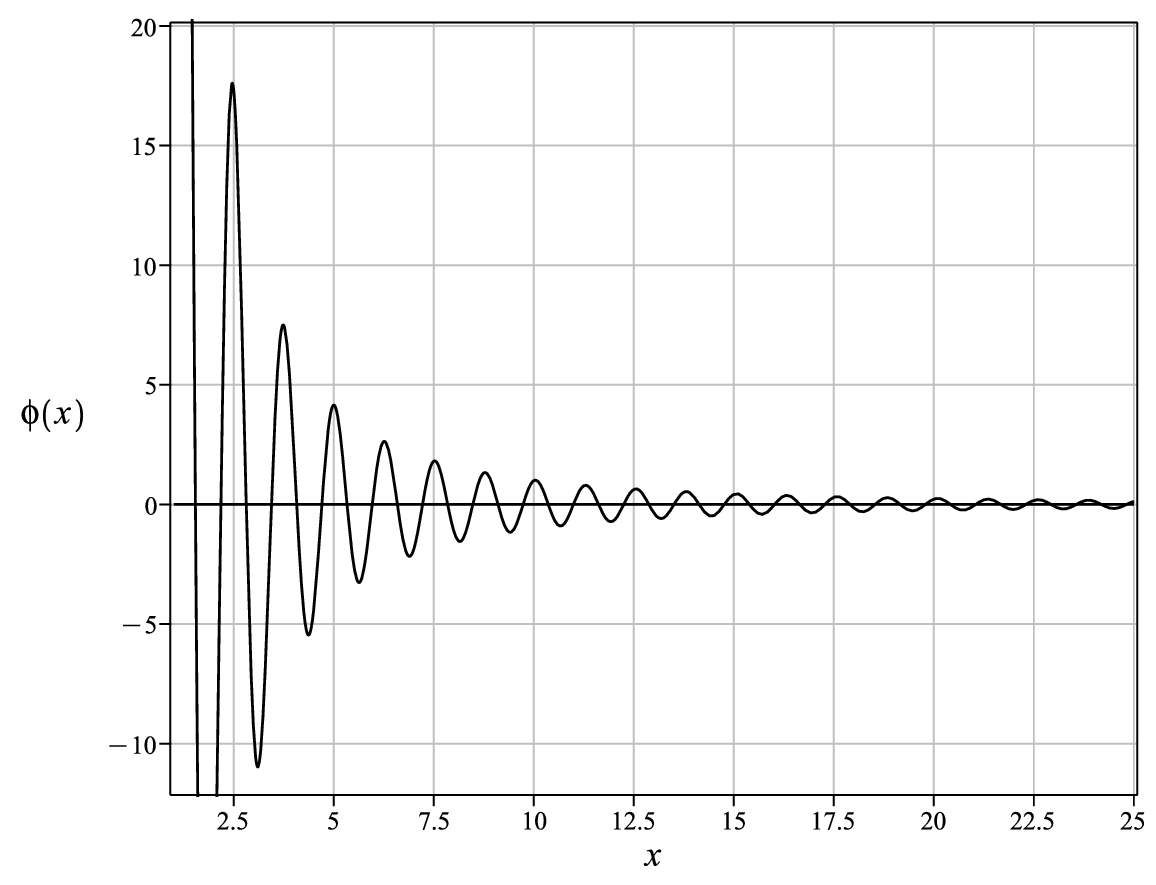}{10}{6.5}{\label{Ignatev4}Damped oscillations of the scalar potential $\phi(x)$ at $\gamma=1$, $\sigma=5$.}

Finally, Figure \ref{Ignatev4} demonstrates damped oscillations of the scalar potential in the case of a bare singularity ($\gamma=1$).

\section*{Conclusion}
Summing up the article, we note its main results.
\begin{itemize}
\item The well-known Fisher solution is reformatted, on the basis of which its exact partial solutions are obtained explicitly and their asymptotic properties are studied.
\item It is shown that in the case of the Higgs interaction potential, the zero approximation of the problem of the smallness of the scalar charge is the vacuum scalar field corresponding to its zero energy and the Schwarzschild gravitational field
- de Sitter with a cosmological constant determined by the square of the vacuum potential.
\item In the first approximation of the smallness of the scalar charge, the scalar field is determined by the field equation against the background of the Schwarzschild - de Sitter metric, while the standard term in the field equation changes sign and doubles.
\item Partial exact solutions of the resulting scalar field equation for perturbations are found and their correspondence with Fisher's solutions is established.
\item The asymptotic behavior of solutions to the resulting field equation near the singularity and horizons, as well as at spatial infinity, has been studied.
\item The oscillatory nature of solutions to the field equation far from the horizons of the Black Hole has been established.
\item The macroscopic average energy densities and pressures of scalar oscillations are calculated and it is shown that the total macroscopic oscillation energy density is negative for $\Lambda>0$ and positive for $\Lambda<0$.
\item It is shown that the macroscopic oscillation energy density leads to a change in the observed mass of the black hole.
\item Based on numerical simulation, the behavior of the scalar field of a Black Hole is demonstrated.
\end{itemize}

Note that in connection with the theory of the formation of supermassive Black holes in the early Universe constructed on the basis of the mechanism of scalar-gravitational instability, the appearance of a scalar halo with negative energy outside the horizons of the Black Hole could become an additional source of information when observing these objects.
\subsection*{Funding}
This paper has been supported by the Kazan Federal University Strategic Academic Leadership Program.

\setcounter{section}{0}
\setcounter{equation}{0}
\setcounter{figure}{0}


\end{document}